\documentclass{article}
\usepackage{epsf,latexsym}

\newcommand{\cqg}{Class. Quantum Grav.}

\newcommand{\jmp}{J. Math. Phys.}
\newcommand{\np}{Nucl. Phys. B}
\newcommand{\pl}{Phys. Lett.}
\newcommand{\pr}{Phys. Rev. D}
\newcommand{\prep}{Phys. Rep.}
\newcommand{\prl}{Phys. Rev. Lett.}

\newcommand{\sjc}[6]{\left\{ \matrix{#1 & #2 & #3 \cr #4 & #5 &#6} \right\}}

\newcommand{\putfig}[2]{$$\leavevmode\hbox{\epsfxsize=#2cm\epsffile{#1.eps}}$$}

\newcommand{\insertfig}[2]{\leavevmode\vcenter{\hbox{\epsfxsize=#2cm\epsffile{#1.eps}}}}

\title{\bf Discrete structures in gravity}
\author{Tullio Regge$^{a)}$ and Ruth M. Williams$^{b)}$  \\                \\
$^{a)}$ Dipartimento di Fisica, Politecnico di Torino,  \\
Corso Duca degli Abruzzi, 10129 Torino, Italy.  \\
$^{b)}$ Girton College, Cambridge CB3 0JG, and   \\
DAMTP, Silver Street, Cambridge CB3 9EW, United Kingdom.}

\begin{document}
\maketitle
%%%%%%%%%%%%%%%%%%%%%%%%%%%%%%%%%%%%%%%%%%%%%%%%%%%%%%%%

\section*{Abstract}
Discrete approaches to gravity, both classical and quantum, are
reviewed briefly, with emphasis on the method using piecewise-linear
spaces.  Models of 3-dimensional quantum gravity involving 6j-symbols
are then described, and progress in generalising these models to four
dimensions is discussed, as is the relationship of these models in
both three and four dimensions to topological theories.  Finally, the
repercussions of the generalisations are explored for the original
formulation of discrete gravity using edge-length variables.

\section*{I Introduction to discrete gravity}
\subsection*{A Basic formalism}
The original motivation for the development of a discrete formalism
for gravity \cite{regge} arose from a number of problems with the continuum
formulation of general relativity.  These included the difficulty of
solving Einstein's equations for general systems without a large
degree of symmetry, the problems of representing complicated
topologies and the need for considerable geometric insight and
capacity for visualisation.  It turned out, as we shall see, that the
discretisation scheme to be described not only helped with these
problems but also found a vital r\^ ole in numerical relativity and in
attempts at a formulation of quantum gravity.

The related branches of mathematics which found their application to
physics in
this formulation of gravity are those of piecewise-linear spaces and
topology and the geometric notion of intrinsic curvature on polyhedra.
The immediate aim was to develop an approach to general relativity
which avoided the use of coordinates, since the physical predictions
of the theory are coordinate-independent.  The basic idea of the
approach, which has subsequently become known as {\it Regge calculus},
is as follows.  Rather than considering spaces (or space-times) with
continuously varying curvature, we deal with spaces where the
curvature is restricted to subspaces of codimension two.  This is
achieved by considering collections of n-dimensional blocks, which are
glued together by identification of their flat (n-1)-dimensional
faces.  The curvature lies on the (n-2)-dimensional subspaces, known
as {\it hinges} or {\it bones}.  For technical reasons, it is
convenient to use blocks which are {\it simplices} (triangles,
tetrahedra and their higher dimensional analogues).

Consider first the realisation of these ideas in two dimensions.  Here
we have examples in everyday life, geodesic domes; these consist of
networks of flat triangles which are fitted together to approximate
curved surfaces, usually parts of a sphere.  Since two triangles with a
common edge can be flattened out without distortion, there is no
curvature on the edges.  However, when a collection of triangles
meeting at a vertex is flattened, there will be a gap, indicating the
presence of curvature at the vertex.  The amount of curvature there
depends simply on the size of the gap or {\it deficit angle}.

It is relatively simple to visualise the generalisation of a
triangulated surface to three dimensions, where a collection of flat
tetrahedra are glued together on their flat triangular faces.  In
general, the tetrahedra at an edge will not fit together exactly in
flat space, so there will be a deficit angle at that edge giving a
measure of the curvature there.  In four dimensions, the curvature is
restricted to the triangles between the tetrahedra where the
four-simplices meet.  And so on in higher dimensions.  Thus we have a
set of flat simplices glued together to approximate a curved space.  

There is another way of viewing the scheme that has just been
described.  Piecewise-flat spaces are interesting in their own right,
so in addition to using them as an approximation scheme for some
curved \lq\lq reality", we may also study such spaces for their own sake.
It has been argued (for example by Friedberg and Lee  \cite{tdlee})
that space-time is actually discrete at the smallest scales, so one
could also regard curved spaces as approximations to a discrete
reality.  \` A  chacun ses go\^ uts!

In order for the piecewise-flat spaces to be of any practical use in
relativity, beyond ease of visualisation, it must be possible to
calculate geometric quantities like curvature and volume, and in
particular to evaluate the Einstein action of such a space.  In
\cite{regge} it
was shown heuristically that the analogue of the Einstein action
\begin{equation}
{I} =
{1 \over 2} {\int R \sqrt g  d^nx},
\end{equation}
is given by 
\begin{equation}
{I_R} =
{\sum_{hinges \ i} |\sigma^i| \epsilon_i}
\end{equation}
where $|\sigma^i|$ is the measure of a hinge $\sigma^i$ and
$\epsilon_i$ is the deficit angle there, equal to $2\pi$ minus the sum
of the dihedral angles between the faces of the simplices meeting at
that hinge.  Rigorous justification for this formula followed in
\cite{cheeger},
where it was shown that it converges to the continuum form of the
action, in the sense of measures, provided that certain conditions on
the fatness of the simplices are satisfied.  Friedberg and Lee
\cite{friedberg} 
approached the problem from the opposite direction, deriving the Regge
action from a sequence of continuum spaces approaching a discrete one.

The reason for choosing the building blocks to be simplices is that
the geometry of a flat simplex is completely determined by the
specification of its edge lengths, so a simplicial space may be
described exactly by these lengths without the need for any further
variables like angles.  This means that the simplest choice of
variables for the discrete theory is the edge lengths; clearly the
action may be calculated once they are specified and they are also the
obvious analogues of the metric tensor, which serves as variable in
the continuum theory.  There, an elegant way of deriving Einstein's
equations is from the principle of stationary action, varying $I$ with
respect to the metric.  The analogue in Regge calculus is to vary
$I_R$ with respect to the edge lengths, giving the simplicial
equivalent of Einstein's equations:
\begin{equation}
{\sum_i { \partial |\sigma^i| \over \partial l_j} \epsilon_i} = 
{0},
\end{equation}
where we have used the result in \cite{regge}, that the variation of the
angular terms gives zero when summed over each simplex
(Schl\" afli's differential identity).  

At first sight, it appears that there is one equation for each
variable, promising the possibility of a complete solution for the
edge lengths.  However the situation is not as simple as that; there
are analogues of the Bianchi identities in Regge calculus
\cite{{regge},{rocek1},{miller},{brewin1},{tuckey1}}, which
in the case of flat space provide exact relations between sets of
equations, and approximate relations in the nearly-flat case, so the
equations may not provide sufficient information for a complete
solution.  In that case there is freedom to specify certain variables,
in analogy with the freedom to specify lapse and shift in the 3+1
version of continuum general relativity.

\subsection*{B Classical applications}

In the ten years after its formulation, Regge calculus was applied
almost exclusively to problems in classical relativity, in particular
to the time development of simple model universes.  (Rather than give a
complete list of references here, we refer the reader to the
bibliography \cite{williams1} which contains a comprehensive list for
the first 20
years.)  The basic idea was really 3+1 in nature: take a triangulation
of a 3-dimensional surface (usually closed but not necessarily so) to
represent a hypersurface at a particular moment of time and join its
vertices to the corresponding vertices of a second 3-dimensional
triangulation, representing the same hypersurface at a later time.
The edges used to join these vertices are taken to be timelike and the
slice of 4-dimensional space-time between the two triangulations is
then divided into 4-simplices by inserting appropriate diagonals.
Given the edge lengths on the first 3-d triangulation, and specifying
the timelike edge lengths, the Regge equations may in principle be
solved for the edge lengths on the second 3-d triangulation. By
repetition of this process, the classical evolution of the inital
spacelike surface may be calculated.  This sounds simple enough, but
unless quite strong assumptions of symmetry are made, the numerical
calculation, involving large sets of simultaneous equations for the
edge lengths, can be very time-consuming and complicated.

Significant progress with this approach was made in the early nineties
when, based on an idea of Sorkin \cite{sorkin}, it was realised that
in general,
the Regge equations decouple into a collection of much smaller
groups.  These groups of equations can then be solved in parallel,
which means that the computer time required for an equivalent
calculation is much less.  This parallelisable implicit evolution
scheme is described in detail in \cite{barrett1} and the basic
mechanism is as
follows.  Consider a single vertex in a triangulated 3-dimensional
spacelike hypersurface and introduce a new vertex \lq\lq above" this.
Connect the new vertex by a \lq\lq vertical" edge to the chosen vertex,
and
by \lq\lq diagonal" edges to all the vertices in the original
hypersurface
to which the chosen vertex was joined.  Each tetrahedron in the
original surface now has based on it a 4-simplex, with apex at the new
vertex.  Note that there is one diagonal corresponding to each edge in
the original vertex radiating from the chosen vertex.  We now use the
Regge equations for these edges in the original surface and for the
vertical edge; the only unknown edges which these equations involve
are the new vertical edge and the diagonal edges, and there is
precisely the same number of equations as unknowns.  Thus, in
principle, we can solve exactly for the unknown edge lengths.  (In
practice,
because of the approximate relationship between the equations from the
Bianchi identities, it is often more convenient to ignore some of the
equations and instead specify conditions equivalent to the lapse and
shift.)  

We have described how to evolve vertices one-by-one in the Sorkin
evolution scheme, and the entire hypersurface can be evolved in this
way.  The method is very general and can be used for a hypersurface
with arbitrary topology.  However, advancing the vertices one-by-one
will not ordinarily be the most efficient way of evolving a
hypersurface.  If any two vertices in a hypersurface are not connected
by an edge, then they can be evolved to the next surface at the same
time without interfering with each other, which is why the method is
obviously parallelisable.

\subsection*{C Some quantum applications}

The earliest application of Regge calculus to quantum gravity was in
three dimensions \cite{ponzano} and involved 6j-symbols.  This work, and
subsequent developments along those lines, will be the subject of the
next two main sections and we shall not discuss it further here.

>From the early eighties onwards, there have been many attempts to
formulate a theory of quantum gravity based on Regge calculus, and we
shall summarise the salient features of some of those approaches, both
analytic and numerical.

The first work on quantum Regge calculus in four dimensions involved
using a study of small perturbations about a flat background to relate
the discrete variables with their continuum counterparts
\cite{rocek2}.  The
discrete propagator was derived in the Euclidean case and shown to
agree with the continuum propagator in the weak field limit. (More
details of this calculation will be given in the section on area Regge
calculus.)  The technique of weak field approximation has proved to be
very useful not only for comparisons with the continuum theory but
also as a guide in numerical calculations.

The difficulties of analytic calculations in quantum Regge calculus,
coupled with the need for a non-perturbative approach and also the
availability of sophisticated techniques developed in lattice gauge
theories, have combined to stimulate numerical work in quantum grvity,
based on Regge calculus.  One approach is to start with a Regge
lattice for, say, flat space, and allow it to evolve using a Monte
Carlo algorithm (see for example \cite{{hamber1},{hamber2},{berg}}).
Random fluctuations are made in the edge lengths
and the new configuration is rejected if it increases the action, and
accepted with a certain probability if it decreases the action.  The
system evolves to some equilibrium configuration, about which it makes
quantum fluctuations, and expectation values of various operators can
be calculated.  It is also possible to study the phase diagram and
search for phase transitions, the nature of which will determine
the vital question of whether or not the theory has a continuum limit.
Many of the simulations have involved an action with an extra term,
quadratic in the curvature, to avoid problems of convergence of the
functional integral; some have included scalar fields
coupled to gravity \cite{hamber3}.  Recent work by Riedler and
collaborators in four dimensions describes evidence for a new
continuous phase transition, essential for a continuum limit, at
negative gravitational coupling \cite{riedler}.

The choice of measure in the functional integral is still a matter for
controversy, depending both on attitude to simplicial diffeomorphisms
and also on the stage at which translation from the continuum to the
discrete takes place.  The numerical simulations just described mainly
use a simple scale invariant measure \cite{hamber4}.  Menotti and
Peirano have derived an expression for the functional measure in
2-dimensional Regge gravity, starting from the DeWitt supermetric,
and giving exact expressions for the Fadeev-Popov determinant for both
$S^2$ and $S^1$x$S^1$ topologies (see \cite{menotti} and references
therein).

A rather different approach to numerical simulations of quantum
gravity is that of {\it dynamical triangulations}. (For a review
containing an extensive set of references, see \cite{ambjorn1}).
This also uses
Regge lattices and the Regge action, but there are important
differences.  In the traditional approach, we are effectively
integrating over the edge lengths in the functional integral, but in
dynamical triangulations, the lattice is taken to be equilateral, with
a certain length scale, and the summation is over different
triangulations, which are generated by a set of {\it (k,l) moves}
\cite{pachner}.  In two
dimensions, there are just two possible moves (and their inverses):
the reconnection of vertices in two triangles with a common edge, and
the insertion of a vertex and edges in a triangle to divide it into three
triangles (2-2 and 1-3 moves).  There are straightforward
generalisations of these moves to higher dimensions.  The moves are
ergodic in the sense that any combinatorially equivalent triangulation
can be generated by a finite succession of these moves.  It is argued
that the restiction to equilateral triangulations is a way of avoiding
over-counting gauge-related configurations.  The approach has been
very successful in two dimensions, where there are analytic results
with which to compare the calculations.  In three and four dimensions,
there has been progress in, for example, deriving the crucially
important exponential bound on the number of triangulations for a
given number of vertices \cite{carfora}, but there are still open
questions on the
continuum limit, since the phase transition appears to be first order
(see the review by Loll \cite{loll}).  Recently a Lorentzian version of
dynamical triangulations has been formulated in (1+1)-dimensions
\cite{ambjorn2}.  Numerical simulations have revealed a new
universality class for
pure gravity, with Haussdorf dimension two.

Discrete gravity has also proved very useful in calculations of the
wave function of the universe \cite{hartle1}.  According to the
Hartle-Hawking
prescription, the wave function for a given 3-geometry is obtained by
a path integral over all 4-geometries which have the given 3-geometry
as a boundary.  To calculate such an object in all its glorious
generality is impossible, but one can hope to capture the essential
features by integrating over those 4-geometries which might, for
whatever reason, dominate the sum over histories.  This has led to the
concept of {\it minisuperspace} models, involving the use of a single
4-geometry (or perhaps several).  In the continuum theory, the
calculation then becomes feasible if the chosen geometry depends only
on a small number of parameters, but anything more complicated soon
becomes extremely difficult.  For this reason, Hartle \cite{hartle2}
introduced
the idea of summing over {\it simplicial} 4-geometries as an
approximation
tool in quantum cosmology.  Although this is an obvious way of
reducing the number of integration variables, there are still
technical difficulties: the unboundedness of the Einstein action
(which persists in the discrete Regge form) leads to convergence
problems for the functional integral, and it is necessary to rotate
the integration contour in the complex plane to give a convergent
result \cite{{hartle3},{louko}}.  

In principle, the sum over 4-geometries should include not only a sum
over metrics but also a sum over manifolds with different topologies.
One then runs into the problem of classifying manifolds in four and
higher dimensions, which led Hartle \cite{hartle4} to suggest a sum
over more
general objects than manifolds, {\it unruly topologies}.  Schleich and
Witt \cite{schleich} have explored the possibility of using conifolds,
which
differ from manifolds at only a finite number of points, and this has
been investigated in some simple cases \cite{{birm1},{silva}}.
However, a sum over
topologies is still very far from implementation.

Yet another area of application of Regge calculus in quantum gravity
involves the study of the simplicial supermetric, the metric on the
space of 3-geometries.  Its signature is crucial for determining
spacelike surfaces in superspace, which are important in Dirac
quantisation and in quantum cosmology.  In the continuum, there are
limited results on the signature and this led to the possibility of
investigating it in the discrete case \cite{hartle5}, where the
analogue is the
Lund-Regge supermetric \cite{lund}.  This supermetric was constructed for
some simple manifolds ($S^3$ and $T^3$) and its signature calculated.
The results agreed with the continuum predictions and also showed that
the supermetric can become degenerate.  We still do not have a
complete understanding of the division of the modes into \lq\lq vertical"
(corresponding to metrics related by diffeomorphisms) and
\lq\lq horizontal" ones.

\subsection*{D Other approaches to discrete gravity}
Of course Regge calculus is not the only way of setting up a theory of
discretised general relativity.  In this subsection, we shall describe
some alternatives.

One important class of schemes involves treating gravity as a gauge
theory.  For example, Mannion and Taylor \cite{mannion} defined a
theory of
gravity on a fixed hypercubic lattice, and Kaku \cite{kaku} used a
fixed
random lattice.  However a dynamical lattice seems more appropriate in
a theory aiming to describe the quantum fluctuations of space-time,
and this was used in much earlier work by Weingarten \cite{weingarten}.

In an approach closely related to Regge calculus, Caselle, D'Adda and
Magnea \cite{caselle} defined a theory of gravity on the dual lattice,
giving
both first- and second-order formulations.  The action they obtained
was a compactified form of the Regge action, involving the sine of the
deficit angle.  D'Adda and Gionti showed \cite{gionti} that Regge
calculus is a
solution of the first order formulation in the limit of small deficit
angles.  The action of Caselle, D'Adda and Magnea was also used by
Kawamoto and Nielsen \cite{kawamoto} in their version of lattice
gravity with fermions.

Immirzi investigated the links between canonical
general relativity in the continuum, loop quantum gravity, and spin
networks, in an attempt to formulate a quantised version of discrete
gravity in the spirit of Regge calculus but ran into problems over
hermiticity \cite{immirzi}. 

A totally different approach to discrete gravity is 't Hooft's polygon
model in (2+1)-dimensions \cite{thooft1}.  This was introduced as a
way to
refute Gott's claim of acausality in (2+1) gravity coupled to point
particles \cite{gott}. 't Hooft's method is to split space-time into
the
direct product of cosmological time and a Cauchy surface tessellated
by flat polygons.  The local flatness of space-time in the pure
gravity regime and the cone-like structure introduced by particles, as
in Regge calculus \cite{rocek3}, are expressed in terms of conditions
on the
edges and vertices of the polygons.  A local Lorentz frame is attached
to each polygon and two constraints imposed; these are firstly that
time runs at the same rate in each polygon (which corresponds to a
partial gauge fixing) and secondly that all vertices are trivalent
(which is acceptable because higher order vertices can always be split
into trivalent vertices connected by edges of zero length).  The
consequences of these conditions are that the length and velocity of
an edge are the same in both polygons to which it belongs, and that
the velocity of each edge is orthogonal to it in both frames.  These
facts result in transition rules for the vertices in the tessellation.

The method for evolving such a space-time is as follows.  Initial data
(lengths and velocities), subject to consistency conditions, are
assigned to the edges on a polygonally-tessellated hypersurface.  The
configuration evolves linearly until an edge collapses to zero length
or a vertex crosses another edge.  A transition, governed by the
vertex conditions, then takes place to another configuration which
will, in general, have different numbers of vertices, edges and
polygons.  The new data will still satisfy the consistency conditions
and the process is then repeated.  When there are particles present at
the vertices, there are deficit angles proportional to their masses,
and the transition rules are modified accordingly.

It is not an easy task to follow the time-evolution of a
(2+1)-dimensional model with particles, even though the system has a
finite number of degrees of freedom.  't Hooft did numerical
simulations on a small computer, with some unexpected predictions.
His big-bang and big-crunch hypotheses were based on the evolution of
a Cauchy surface with $S^2$ or $S^1$x$S^1$ topology, tessellated by a
single polygon \cite{thooft2}.  It would be interesting to test these
predictions for more complex initial configurations, and as a means to
this, there has been recent work \cite{hollmann} in which the
constraint
equations have been interpreted in terms of hyperbolic geometry (see
also \cite{thooft3}), and various consistent sets of initial data set
up, but
the evolution calculations have not yet been completed.  Part of the
motivation for this work is to compare 't Hooft's method with other
approaches to (2+1) gravity, in particular Regge calculus.  A
(2+1)-dimensional code has been set up for Regge space-times and a
number of calculations performed \cite{gentle}, with a view to making
detailed
comparisons with the 't Hooft method.  The ultimate aim is to
understand the exact relationship between the two approaches, which
seem rather different but have many concepts in common.  

Based on the polygon approach, various toy models of (2+1)-dimensional
gravity have been constructed \cite{{thooft2},{welling}}, issues of
topology been
addressed \cite{{welling},{franzosi1}} and particle decay and
space-time kinematics been
investigated \cite{franzosi2}.  't Hooft himself has
proposed quantised models
of (2+1)-dimensional space-time \cite{thooft5}, showing that
gravitating particles live on a space-time lattice.  For an
$S^2$x$S^1$ topology, first quantisation of Dirac particles is
possible.  Waelbroeck has suggested a similar approach, using canonical
quantisation in (2+1) dimensions \cite{waelbroeck}. 

Back at the classical level, Brewin has formulated \cite{brewin2} a
discretisation of gravity which he feels is closer to the original
theory of general relativity.  Preliminary calculations are
encouraging.  For other important work on lattice gravity
by Bander, Jevicki and Ninomiya, Khatsymovsky and Lehto, Nielsen and
Ninomiya, we refer the reader to the Regge calculus review and
bibliography \cite{williams1}.  We emphasise again that this paper
is not meant to
be an exhaustive review of the subject.

After this rather rapid survey of applications of Regge calculus, and
some other approaches to discrete gravity, we
shall now concentrate on one particular approach and show how it has
led to exciting new developments in the search for a quantum theory of
gravity.

\section*{2. 6j-symbols in 3-dimensional quantum gravity}

As promised, we now look in detail at the earliest link forged between
Regge calculus and quantum gravity, now known as the Ponzano-Regge
model \cite{ponzano}.  This emerged from a paper on 6j-symbols and
we will first
give the background to these.

\subsection*{A 6j-symbols}

6j-symbols, which are generalisations of the more well-known
Clebsch-Gordan coefficients, first arose as tools for the computation
of matrix elements in the theory of complex spectra \cite{racah}, and
are now
used routinely by atomic physicists and theoretical chemists 
in quantum mechanical calculations involving angular momentum.  In
particular, they relate the possible basis wave functions when three
angular momentum are added:

\putfig{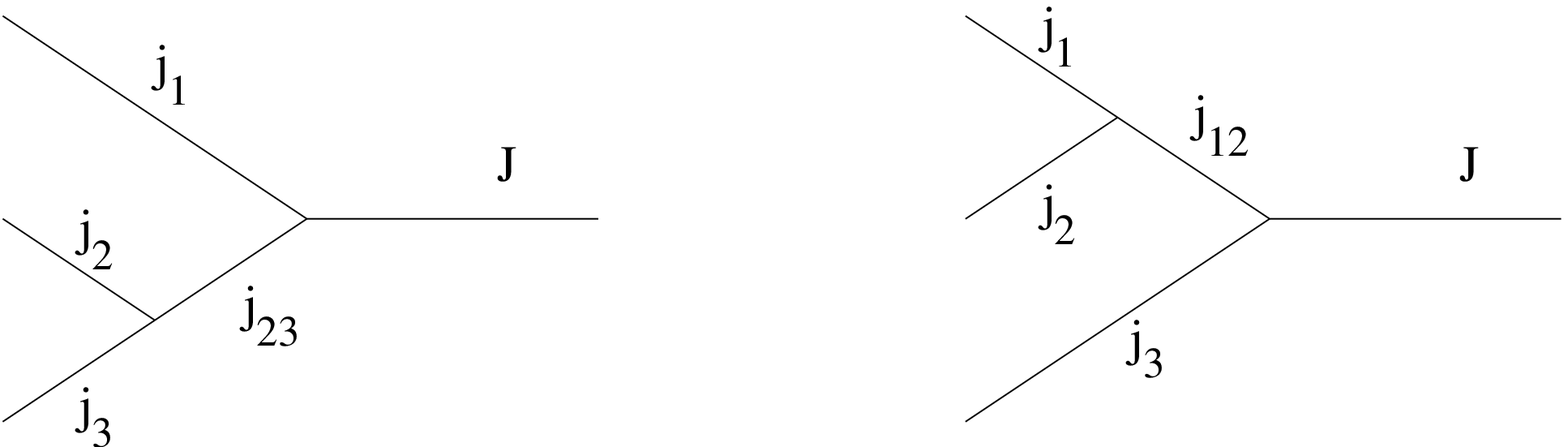}{8}

\[
{|j_1,(j_2,j_3)j_{23},J>} =
\]
\begin{equation}
{\sum_{j_{12}} \sqrt{(2j_{12} + 1)(2j_{23} + 1)} (-1)^{j_1 + j_2 +
j_3 + J} \sjc{j_1}{j_2}{j_3}{j_{12}}{j_{23}}{J} |(j_1,j_2)j_{12},j_3,J>}.
\end{equation}

%\[ \sjc{j_1}{j_2}{j_3}{j_{12}}{j_{23}}{J} \]

An alternative and useful definition involves the {\it recoupling}
diagram:

\[
\insertfig{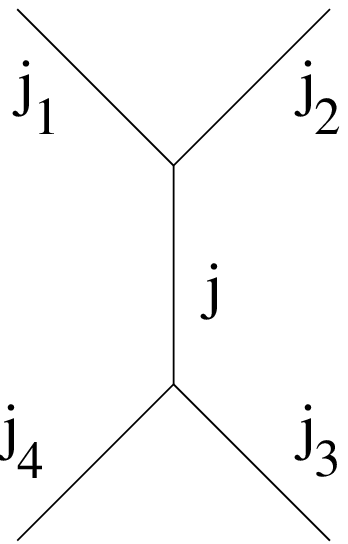}{1.5} = \sum_{J} \sjc{j_1}{j_2}{j}{j_3}{j_4}{J} \insertfig{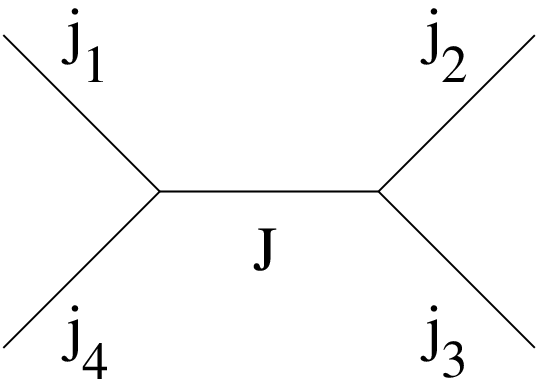}{2.5}
\]

A graphical representation is obtained by associating a 6j-symbol with
a tetrahedron:

\begin{equation}
\insertfig{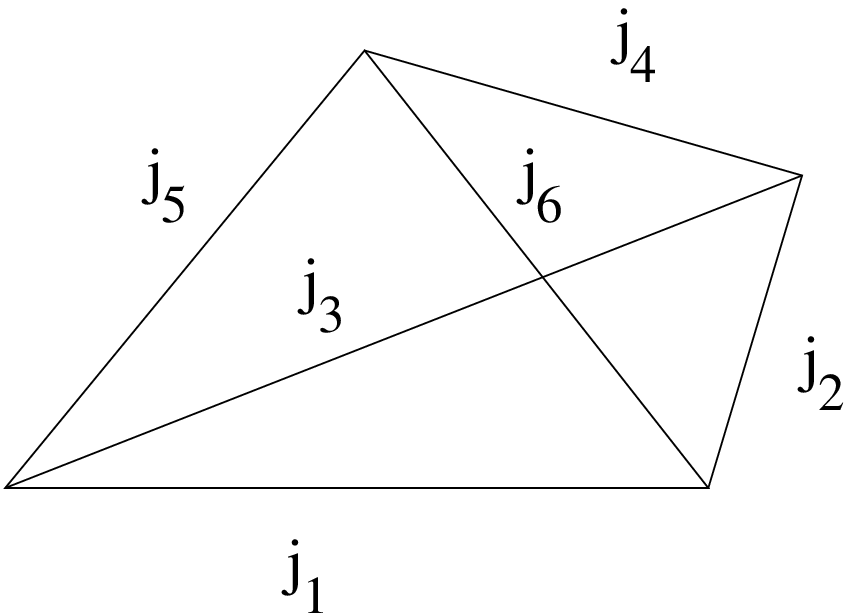}{4}
\end{equation}
with the arguments of the 6j-symbol corresponding to the edge lengths
of the tetrahedron.  (For technical reasons, it turns out to be more
accurate to associate a symbol with arguments $a,b,...$ to a
tetrahedron with edge lengths $a + 1/2, b + 1/2,...$.)  For the
6j-symbol to be non-zero, the arguments have to satisfy the analogue
of the triangle inequalities for each face of the tetrahedron:

\begin{equation}
{j_3 \leq j_1 +j_2\ \ \ \ \  etc}
\end{equation}

They can be evaluated from the formula

\[
\sjc{a}{b}{c}{d}{e}{f} =
\sqrt{\Delta (a,b,c) \Delta (a,e,f) \Delta (c,d,e) \Delta (b,d,f)}
\]
\[
\sum_x (-1)^x (x + 1)![(a+b+d+e-x)!(a+c+d+f-x)!(b+c+e+f-x)!
\]
\begin{equation}
(x-a-b-c)!(x-a-e-f)!(x-c-d-e)!(x-b-d-f)!]^{-1}
\end{equation}

where 

\begin{equation}
{\Delta (a,b,c)} =
{(a+b-c)!(b+c-a)!(c+a-b)![(a+b+c+1)!]^{-1}}.
\end{equation}

These 6j-symbols are based on the group $SU(2)$, but as we shall see,
it is also possible to have q-deformed 6j-symbols based on quantum
groups.  For example, define \cite{kirillov}

\begin{equation}
{q = exp(2\pi i/r)}
\end{equation}

and 

\begin{equation}
{[n]} =
{(q^{n/2} - q^{-n/2}) \over (q^{1/2} - q^{-1/2})}.
\end{equation}

Then the q-deformed 6j-symbol for $SU_q(2)$ is defined in the same way
as the undeformed one, with $n$ replaced everywhere by $[n]$.  Note
that $[n] \rightarrow n$ as $q \rightarrow 1$ and $r \rightarrow
\infty$.

\subsection*{B The Ponzano-Regge model}
The main purpose of the paper by Ponzano and Regge \cite{ponzano} was
to derive
asymptotic formulae for classical (ie undeformed) 6j-symbols in the
limit when certain arguments became large.  The case most relevant to
the exposition here is when all six parameters become large.  The edge
lengths of the corresponding tetrahedron are really related to $j_i
\hbar$ and these quantities are kept finite as $j_i \rightarrow
\infty$ while $\hbar \rightarrow 0$ so this process corresponds to the
semi-classical limit.  This asymptotic behaviour is given by

\begin{equation}
{\sjc{j_1}{j_2}{j_3}{j_4}{j_5}{j_6} \sim}
\frac{1}{12 \pi V} cos (\sum_i j_i \theta _i + {\pi / 4})
\end{equation}

where $V$ is the volume of the terahedron and $\theta _i$ is the
exterior dihedral angle at edge $i$ (ie the angle between the outward
normals to the faces meeting there).  This was recently proved
rigorously by Roberts \cite{roberts1}. 

To see the connection between this formula and quantum gravity,
consider the following {\it state sum} defined in \cite{ponzano}.
Take a closed
2-dimensional surface, triangulate it and divide its interior into
tetrahedra, possibly inserting internal vertices.  Label the internal
edges by ${x_i}$ and the external ones by ${l_i}$.  Define 

\begin{equation}
{S(\{l_i\})} =
{\sum _{{x_i}} \prod _{tetrahedra} {\{6j\}} (-1)^X \prod _i (2x_i +
1)}
\end{equation}

where the $X$ in the phase factor is a function of the edge lengths. 

Although this expression is infinite in many cases, it has some
extremely interesting properties.  In particular, noting that in the
sum over the internal edges, the large values dominate, we can replace
the sum by an integral with respect to those edge lengths and use the
asymptotic formula stated above.  Then the dominant contribution to
the integral comes from the points of stationary phase, which are
given by

\begin{equation}
{\sum _{tetrahedra\ k\ meeting\ on\ edge\ i} (\pi - \theta ^k_i)} =
{2 \pi}.
\end{equation}

This means that the sum of the dihedral angles at each edge is $2
\pi$, which is precisely the condition for local flatness in a
3-dimensional simplicial space.  What is more, the state sum is given
approximately by

\begin{equation}
{S \approx}
\frac{1}{\sqrt{12 \pi}} \int \prod _i dx_i (2x_i + 1) (-1)^X \prod
_{tet\ k} \frac{1}{\sqrt{V_k}} cos (\sum _{l\in tet\ k} j_l
\theta^k _l + {\pi \over 4}).
\end{equation}

Now this contains a term of the form 

\[
{\int \prod_i dx_i (2x_i + 1) exp(i\sum _{edges\ l} j_l (2 \pi - \sum
_{tet\ k \ni i} (\pi - \theta ^k_l))} = 
\]
\begin{equation}
{\int \prod _i dx_i (2x_i + 1) exp(i\sum j_l \epsilon _l)},
\end{equation}

which looks precisely like a Feynman sum over histories with the Regge
calculus action in three dimensions:

\begin{equation}
{\int \prod _i d\mu (x_i) exp(iI_R)}
\end{equation} 

with 

\begin{equation}
{I_R} =
{\sum _l j_l \epsilon _l},
\end{equation}

where $\epsilon _l$ is the deficit angle at edge $l$ and $d\mu(x_i)$
is the measure on the space of edge lengths. 

This result was rather puzzling and, although Hasslacher and Perry
\cite{hasslacher} emphasised the connection between spin networks
and simplicial
gravity, its significance was not fully
appreciated until much later, when a very similar expression was
written down in a different context.

\subsection*{C The Turaev-Viro model}

In the late eighties and early nineties, mathematicians put a lot of
effort into searching for invariants of manifolds, the hope being, at
least in part, that such quantities would help with the classification
of manifolds.  Without being aware of the Ponzano-Regge work, Turaev
and Viro \cite{turaev1} defined a state sum for triangulated
3-manifolds, which
in many aspects was identical to that of Ponzano and Regge.  The main
differences were that they gave formulae for closed manifolds as well
as those with boundary, they showed explicitly that the quantity
obtained was independent of triangulation, and finally, they used
6j-symbols for the quantum group $SL_q(2)$.  Only some of the
irreducible representations of this group, the ones with $j$ taking
finite values, have suitable algebraic properties, which means that
the edge lengths are not summed up to
infinite values; $j_i$ can take only integer and half-integer values
from the set $(0,1/2,1,...,(r-2)/2)$, with $r \geq 3$.  A very
important consequence of this is that the answer obtained is finite,
and so the model appears to be a regularised version of the
Ponzano-Regge model.

The obvious question to ask is how the Turaev-Viro state sum is
connected to quantum gravity.  Witten \cite{witten} conjectured
that it was
equivalent to a Feynman path integral with the Chern-Simons action for
$SU_k(2) \bigotimes SU_{-k}(2)$, and this and equivalent results were
proved by a number of people \cite{{turaev2},{walker},{roberts2}}.
To see how this works \cite{{ooguri1},{williams2}},
consider the Chern-Simons Lagrangian for this group product:

\[
{\mathcal{L}} =
{{k\over 4\pi} \int _M Tr(A_+\wedge dA_+ + {2\over3} A_+\wedge
A_+\wedge  A_+)}
\]
\begin{equation}
{- {k\over 4\pi} \int _M Tr(A_-\wedge dA_- + {2\over3} A_-\wedge
A_-\wedge A_-)}
\end{equation}

where 

\begin{equation}
{A_{\pm}} =
{A_{i \ (\pm)}^a T_a dx_i}
\end{equation}

with $T_a$ a basis of the $SU(2)$ Lie algebra.  Making the change of
variables

\begin{equation}
{A_{i \ (\pm)}^a} =
{\omega_i^a \pm {1\over k} e_i^a},
\end{equation}
%\begin{equation}
%{B{_i}{^a}} =
%{\omega{_i}{^a} - {1 \over k} e{_i}{^a}},
%\end{equation}

where $e_i^a$ is the dreibein and

\begin{equation}
{\omega_i^a} =
{{1\over 2} \epsilon^{abc} \omega _{ibc}},
\end{equation}

with $\omega _{ibc}$ being the connection 2-form, we obtain 

\begin{equation}
{\int (e \wedge R + {\lambda _k \over 3} e\wedge e\wedge e)},
\end{equation}

which is the Einstein-Hilbert action for gravity with cosmological
constant given by

\begin{equation}
{\lambda _k} =
{(\frac {4 \pi} {k})^2}.
\end{equation}

(Note that the $k$ here is equal to $r -2$, where $r$ appears in the
definition of $q$.)  By taking the limit as $k \rightarrow \infty$, we
obtain 3-dimensional gravity with zero cosmological constant ie the
theory represented by the Ponzano-Regge model.  This result is
consistent with the fact that the $q \rightarrow 1$ limit of the
Tuaev-Viro model is the Ponzano-Regge model.

The properties of the Turaev-Viro state sum show that the formalism is
an example of a {\it topological quantum field theory} (see eg
\cite{birm2}).
This is
perfectly appropriate for a theory of gravity in three dimensions
where there are no local degrees of freedom.  As for the Ponzano-Regge
theory, the dominant classical configurations are locally flat (recall
that in Chern-Simons gravity, the solutions involve the space of flat
connections.) 

The relationship between the Turaev-Viro invariant and 3-dimensional
quantum gravity is an extremely important one.  It means that in three
dimensions, we have in principle a way of calculating the partition
function for triangulated manifolds.  This has been done for many of
the simpler 3-manifolds (see \cite{{kauffman1},{ionicioiu1}} for
example).  The Turaev-Viro
expression can also be used for calculating topology-changing
amplitudes in 3-dimensional gravity; the method here is to construct a
cobordism between two 2-dimensional triangulated surfaces and then use
the Turaev-Viro expression for a manifold with boundaries to evaluate
the transition probability \cite{ionicioiu2}.

\subsection*{D Spin networks}

The Turaev-Viro expression is not the only method of calculating this
particular invariant of 3-manifolds.  Various other prescriptions have
been written down, and one that is worth describing at this stage is
that using spin networks.  These were invented by Penrose
\cite{penrose} who
wanted to formulate a purely combinatorial approach to space-time.
His networks had trivalent vertices and the edges of the graphs were
labelled by spins.  He developed a method of calculating the value of
an arbitrary spin network and was able to show that this led to the
usual angles of 3-dimensional space.

Penrose's spin networks were later generalised in a number of ways.
The edges were labelled by representations of quantum groups and it
was necessary to introduce intertwining operators or {\it
intertwiners} at the vertices \cite{barrett8}.  In
some cases a framing was introduced and the graphs became "ribbon
graphs" \cite{turaev3}.  Kauffman \cite{kauffman2} showed how to
calculate the
Turaev-Viro invariant by taking the graph dual to a triangulation to
be a spin network; the edges of the graph inherit the labels of the
triangulation edges which they cross.  Spin networks
have also been introduced into loop quantum gravity \cite{rovelli1},
where they
are an important calculational tool, for instance in the
derivation of the spectrum of the area and volume operators
\cite{rovelli1}.  (Note that Freidel and Krasnov also obtained a
discrete spectrum for the volume operator in BF theory by
differentiating the Turaev-Viro amplitude with respect to the
cosmological constant \cite{freidel1}.)  As we shall see, spin
networks also play
a r\^ ole in recent attempts at formulations of 4-dimensional quantum
gravity.

\section*{III Extensions to four dimensions}

After it was realised that the Turaev-Viro state sum provides a finite
theory of 3-dimensional gravity, the search began for a generalisation
to four dimensions.  Before this is described, we shall stop to ask
what we hope to achieve by this.  In classical general relativity,
there are enormous qualitative differences between gravity in three
and four dimensions.  In particular, there are gravitons in four
dimensions, but not in three, so although it seems reasonable to
describe 3-dimensional gravity by a topological invariant of a
manifold, it seems {\it likely} that an invariant of a 4-manifold might
describe only some topological sector of gravity.  We shall return to
this point later.

The obvious way of setting about extending the 3-dimensional model,
based on 6j-symbols, to four dimensions is by using some 3nj-symbol
for a value of $n$ larger than 2.  The 3nj-symbols in the state sum
would then be expanded in terms of 6j-symbols, and the Ponzano-Regge
formula for their asymptotic values inserted, the hope being that this
would give an expression looking like a path integral with the
4-dimensional Regge action.  The problem with this is that the
asymptotic formula involves the 3-dimensional dihedral angles and it
is very difficult to relate these to 4-dimensional angles.  This
indicates that a more radical generalisation may be needed.

We shall now describe some of the attempts at generalisation, leading
up to some recent work which seems very promising.

\subsection*{A The Ooguri model}

A source of inspiration for some generalisations of the Ponzano-Regge
and Turaev-Viro models was Boulatov's generalised matrix model
\cite{boulatov},
which involved a scheme for generating 3-dimensional simplicial
complexes as terms in a perturbative expansion.  The contribution from
each simplicial-complex was weighted by its Ponzano-Regge or
Turaev-Viro invariant, depending on the value of $q$.  Boulatov's
model was formulated in a way that it could be extended to higher
dimensions, and the 4-dimensional case for $q = 1$ was worked out by
Ooguri \cite{ooguri2}.

The essential ingredients in Ooguri's model are the assigning of
group variables to the tetrahedra and spin j labels to the triangles in
the triangulated 4-manifold.  The terms in Ooguri's action are of two
types: the first is a product of two functions of the group variables,
and this represents two glued tetrahedra; the second is a product of
five functions and represents the tetrahedra in a 4-simplex.  A
Fourier decomposition is performed in terms of rotation matrices and
the group variables are then integrated out, using the standard
relationship between rotation matrices and 3j-symbols, and the
invariant Haar measure normalised to unity.  The
resulting expression has four 3j-symbols associated to each
tetrahedron; these may then be divided between the 4-simplices meeting
on that tetrahedron, and then each 4-simplices ends up with ten
3j-symbols which can be combined to give a 15j-symbol.  At first
sight, it seems odd to associate a 15j-symbol with a 4-simplex which
has only 10 triangles labelled by spin values.  The way to interpret
the symbol is to consider the dual graph, which has ten edges and five
4-valent vertices (corresponding to each tetrahedon in the original
triangulation). Each of these 4-valent vertices can be split into two
trivalent ones, and an extra spin label can be assigned to the edge
joining them.  This splitting sounds rather arbitrary but different
splittings are related by 6j-symbols (see the second diagram in the
section on 6j-symbols) and when all summations are performed, the
result is independent of splitting.

The partition function is calculated by integrating the exponential of
minus the action over the Fourier coefficients, and the resulting
expression is  

\begin{equation}
{Z} =
{\sum _C \frac {1}{N{_{symm}}(C)} \lambda ^{N_4(C)} Z(C),}
\end{equation}

with $Z(C)$ given by

\begin{equation}
{Z(C)} =
{\sum _j \prod _{triangles} (2j_t + 1) \prod
_{tetrahedra} \{6j\} \prod _{4-simplices} \{15j\}.}
\end{equation}

The summation in $Z$ is over simplicial complexes $C$, with $N_{symm}$
being the rank of the symmetry group of $C$, and $N_4$ the number of
4-simplices in $C$.  By writing the contributions from all the
tetrahedra meeting on a particular triangle in terms of rotation
matrices, one can show that the holonomy around any triangle is
trivial.  This ties up with the proposed link between Ooguri's model
and BF theory, as we shall see later.

\subsection*{B The Archer, Crane-Yetter and Roberts models}

The extension of the Ooguri model to general values of $q$ was worked
out by various people.  Archer \cite{archer} showed how to construct a
q-deformed topological quantum field theory in general dimension,
giving realisations in three and four dimensions based on the quantum
group $U_q(SLN)$, and suggesting that his theory corresponded to BF
theory with a cosmological constant.

Crane and Yetter \cite{crane1} outlined the construction of a q-deformed
version of Ooguri's model and recognised its relationship with the
work of Roberts \cite{roberts2}, who had defined a 4-dimensional
generalisation of his own \lq\lq chain-mail" formulation of the
Turaev-Viro
invariant.  Roberts showed that his invariant for a 4-manifild $M$
depended on two simple functions of $r$, one raised to the power of
$\sigma (M)$, the signature, and the other to the power of $\chi (M)$,
the Euler character.

The result of Roberts was disappointing but instructive for those
trying to construct a theory of 4-dimensional quantum gravity by this
method.  Since the models do not give any new information about
4-manifolds, it showed that a more radical generalisation was needed. 

\subsection*{C The Barrett-Crane model}

An important step forward in these generalisation attempts has been
taken recently with the formulation of the Barrett-Crane model.
(Although the details of some aspects of the model, and other related
models, have yet to be worked out, we consider the ideas sufficiently
important to include in this review.)  First came the realisation that
it made sense to generalise spin networks to {\it relativistic spin
networks} appropriate to four dimensions \cite{barrett2}.  The
symmetry group
$SO(3)$ in three dimensions is replaced by $SO(4)$ in four dimensions,
which has spin covering $SU(2) \bigotimes SU(2)$.  Barrett and Crane
therefore
label the triangles by two spin labels rather than one.  Thus in a
relativistic spin network, the edges (dual to the triangles in the
4-complex) carry labels $(j_1,j_2)$ and the vertices (dual to
tetrahedra) carry the appropriate intertwiners.  Barrett and Crane
suggested that the two labels $j_1$ and $j_2$ should be equal to
satisfy the constraints at the vertices, and Reisenberger
\cite{reisenberger1} showed that this solution is unique.
Thus the Barrett-Crane model is a constrained
doubling of the earlier attempts described in the previous
subsections, which can thus be regarded as just describing the
self-dual section of gravity.

We now describe the Barrett-Crane model in a little more detail.
Consider a single 4-simplex, draw its dual graph and then split the
vertices as described for the Ooguri model.  The first expression
written down by Barrett and Crane for the amplitude of a 4-simplex was
of the form

\begin{equation}
{I_1} =
{\sum_{extra \ edges} c_j {\{15j\}}^2,}
\end{equation}

where $c_j$ is a weight factor and the 15j-symbol is squared because
of the $(j,j)$ labelling on each edge of the dual graph.  It turned
out to be very difficult to evaluate the asymptotic value of this
expression, so Barrett and Crane tried a second approach.

Label the five tetrahedra in a 4-simplex by $k$; the spin label on the
triangle where tetrahedra $k$ and $l$ meet is then denoted by
$j_{kl}$.  The matrix representing the element $g \in SU(2)$ in the
irreducible representation of spin $j_{kl}$ is denoted by
$\rho_{kl}(g)$.
Variables $h_k \in SU(2)$ are assigned to the tetrahedra and the
invariant $I_2$ (the second Barrett-Crane model) is obtained by
integrating a function of these variables over each copy of $SU(2)$:

\begin{equation}
{I_2} =
{(-1)^{\sum_{k<l} 2j_{kl}} \int _{h \in SU(2)^5} \prod_{k<l} Tr
\rho_{kl}(h_k h_l^{-1}).}
\end{equation}

The measure used is the Haar measure normalised to unity.

The next step is to relate this expression to the geometry of the
4-simplex \cite{barrett3}.  Using the fact that $SU(2)$ is isomorhic
to $S^3$,
and embedding $S^3$ in $R^4$, we can regard the element $h_k \in
SU(2)$ as a unit vector in $R^4$, normal to the 3-dimensional
hyperplane in which tetrahedron $k$ lies.  Then according to a
well-known formula in representation theory,

\begin{equation}
{Tr \rho (h_k h_l^{-1})} =
{\frac {sin (2j+1)\phi} {sin \phi}}
\end{equation}

where $cos \phi = h_k . h_l$ ie $\phi$ is the angle between
the normals and thus the exterior angle between the two hyperplanes. 

Note that the five hyperplanes define a 4-simplex up to translation
and an overall scale.  Thus integration over the elements $h_k$ may be
interpreted as integration over all possible 4-simplices.

Recalling the equivalence of the asymptotic value of the Ponzano-Regge
model to a path integral with the 3-dimensional Regge calculus
action, we now look for a similar result here \cite{barrett4}.  We
write $sin
(2j+1)\phi$ in terms of exponentials and, for large $j$, use the method
of stationary phase to find the asymptotic value of the integral.
Setting $\epsilon _{kl} = \pm 1$, we write $I_2$ as 

\begin{equation}
{I_2} =
{{\frac {(-1)^{\sum_{k<l} 2j_{kl}}} {(2i)^{10}}}
\sum_{\epsilon_{kl} = \pm 1} \prod_{h \in SU(2)^5} {\frac
{\epsilon_{kl}} {sin \phi_{kl}}} exp (i \sum_{s \le l} \epsilon_{kl} (2j_{kl}+1)\phi_{kl}),}
\end{equation}

which makes it clear that we need the stationary points of 

\begin{equation}
{I} =
{\sum_{k<l} \epsilon_{kl} (2j_{kl}+1)\phi_{kl}.}
\end{equation}

Now the $\phi_{kl}$'s for a 4-simplex are not independent variables;
as is shown in the original formulation of Regge calculus
\cite{regge}, their
variations are related by

\begin{equation}
{\sum_{k<l} A_{kl} d\phi_{kl}} =
{0.}
\end{equation}

Adding this constraint to $I$ with a Lagrange multiplier $\mu$, we
find that for each triangle,

\begin{equation}
{\epsilon_{kl} (2j_{kl}+1)} =
{\mu A_{kl}.}
\end{equation}

The overall scale can then be fixed by taking $\mu = \pm 1$.

What has been established is that for a stationary phase point, then
firstly, the angles $\phi_{kl}$ are those of a geometric 4-simplex
with triangle areas

\begin{equation}
{A_{kl}} =
{2j_{kl}+1,}
\end{equation}

and secondly, the integrand is $exp(i\mu I_R)$, with 

\begin{equation}
{I_R} =
{\sum_{triangles \ kl} A_{kl} \phi_{kl},}
\end{equation}

the Regge calculus version of the Einstein action for a 4-simplex,
with $\mu = \pm 1$. 

The formulation of this model is by no means complete.  The next step
is to sum over 4-simplices, which is likely to be more difficult than
for the first model, where the extra labels on tetrahedra  could
provide links between neighbouring 4-simplices.  The resulting
expression will need to be regularised by passing to representations
of the quantum group $U_q(SL2)$, as in the transition from the
Ponzano-Regge state sum to that of Turaev and Viro.  This analogy is
not precise because the Barrett-Crane amplitude is {\it not}
independent of triangulation.  The covariance lost here may perhaps be
restored by summing over triangulations using a generalised matrix
model approach, as suggested by De Pietri, Freidel, Krasnov and
Rovelli \cite{depietri}.  (Note that these authors refer to what we
have called the \lq\lq second Barrett-Crane model" as their \lq\lq
first
version".)

We shall return to the interpretation of this model in the next
subsection, but first note that the formulation described so far is
Euclidean.  There have been Lorentzian models proposed recently: in
(2+1) dimensions, Freidel \cite{freidel2} has set up a
version in which $SU(2)$ is replaced by $SL(2,R)$, for which both
discrete and continuous representations are used.  This results in a
model in which time is discrete and space continuous.  The partition
function requires summation over causal structures, which obviously
has no analogue in the Euclidean case.  The 6j-symbols for the discrete
series representation of $SL(2,R)$ were defined first by Davids
\cite{davids}, who also obtained the analogous Ponzano-Regge formula,
which here involves $exp(iI_L)$, where $I_L$ is the Lorentzian Regge
action.  In (3+1) dimensions, Barrett
and Crane \cite{barrett5} have proposed versions based on the
classical Lorentz group and on the quantum Lorentz algebra, but the
second of these is still at a preliminary stage.

\subsection*{D Relation to BF theory}

This is not the place for a review of BF theory, but let us briefly
mention its relevant properties.  It is a gauge theory which can be
defined in any dimension and is \lq\lq background-free" in the sense
that no
pre-existing metric or other geometrical structure on space-time is
needed.  It is a theory with no local degrees of freedom.

The action for BF theory in four dimensions is

\begin{equation}
{I_{BF}} =
{\int_M Tr(B \wedge F),}
\end{equation}

where $B$ is a Lie algebra-valued 2-form, and $F = dA + A \wedge A$,
with $A$ the connection 1-form.  It gives rise to the constraint $F =
0$, which means that the connection $A$ is flat.  This ties up with
the trivial holonomy around triangles in the Ooguri model.  The other
constraint, $d_AB = 0$, is the statement of a particular type of gauge
symmetry in BF theory.

To understand the relationship between general relativity and BF
theory in four dimensions \cite{baez1}, consider the Palatini
formulation of general relativity, which has action 

\begin{equation}
{I_P} = 
{\int_M Tr(e \wedge e \wedge F),}
\end{equation}

with $e$ a 1-form on the manifold $M$, and $F$ defined in terms of the
connection as for BF theories.  It is immediately apparent that there
is a relationship between this Palatini formulation, and BF theory
with $B$ constrained to be of the form $e \wedge e$.  There is a
subtle difference between the equations of motion derived from the two
actions: for general relativity, we have

\begin{equation}
{e \wedge F = 0 , \ \ \ d_AB = 0}
\end{equation}

as compared with the BF equations

\begin{equation}
{F = 0 , \ \ \ d_AB = 0.}
\end{equation}

Thus the equations of general relativity are weaker here than those
for BF theory, which, heuristically, is why general relativity in four
dimensions is more general than a topological theory.  We see that
general relativity in four dimensions is equivalent to BF theory with
an extra constraint $(B = e \wedge e)$ (giving rise to the paradoxical
statement that adding a constraint produces a less restricted theory!)

We see now a further justification of why, in the Barrett-Crane model,
the two spin labels on each triangle should be equal (ie we see the
parallel between $(j,j)$ and $e \wedge e$).  Thus the constraint
which Reisenberger \cite{reisenberger1} derived may be interpreted as
equivalent to the constraint which relates BF theory to general
relativity in four dimensions.

Reisenberger \cite{reisenberger3} has explored further the
relationship between the Barrett-Crane model and continuum theories,
showing that the model corresponds to an $SO(4)$ BF theory in which
the right- and left-handed areas, defined by the self-dual and
anti-self-dual components of $B$, are constrained to be equal.

Before considering an extension of BF theory in four dimensions, let
us return to the case of three dimensions.  It can be shown that
3-dimensional general relativity without matter is a special case of
BF theory, where the equations of motion give simply that the
connection is torsion-free and flat. 
Adding an extra term to the BF Lagrangian has a very interesting
effect.  Starting from the modified action

\begin{equation}
{I_{BF}'} =
{\int_M Tr(B \wedge F + {\frac {\lambda} {6}} B \wedge B \wedge B)}
\end{equation}

and making the transformation

\begin{equation}
{A_{\pm}} =
{A \pm \sqrt{\lambda} B,}
\end{equation}

we can show that $I_{BF}'$ is equal to the difference of the two
Chern-Simons actions as in section 2.  It was shown there that this
was equivalent to 3-dimensional general relativity with a cosmological
constant $\lambda$ related to the deformation parameter $q$, which
gives a finite theory of quantum gravity in that dimension
\cite{barrett6}.  Thus a r\^ ole of the cosmological constant is to
regularise the theory.

In four dimensions, the extra term that we need seems to be slightly
different.  The proposed modified action is 

\begin{equation}
{I_{BF}''} =
{\int_M Tr(B \wedge F + {\frac {\lambda} {12}} B \wedge B).}
\end{equation}

The form of this extra term was first suggested by Archer
\cite{archer}, whose contribution is described earlier.  It has been
discussed more recently by Baez \cite{baez1,baez2}, who gives a very
comprehensive discussion of BF theory and the discrete models of
quantum gravity in three and four dimensions.  (Reference \cite{baez1}
is recommended strongly for fuller details of these issues.)  Imposing
the constraint $B = e \wedge e$ as before, the action becomes that for
the Palatini formulation of general relativity with cosmological
constant, 

\begin{equation}
{I_P'} =
{\int_M Tr(e \wedge e \wedge F + {\frac {\lambda} {12}} e \wedge e
\wedge e \wedge e).}
\end{equation}

This suggests the possibility of finding a regularised version of
4-dimensional quantum gravity by constructing a q-deformed version of
the Barrett-Crane model, satisfying the relationship 

\begin{equation}
{\lambda \rightarrow 0 \ as \ q \rightarrow 1.}
\end{equation}

Another possible (and related) way forward is through spin foam
models, as described briefly in the next subsection.

\subsection*{E Spin foam}

As mentioned in the section on 3-dimensional gravity, spin networks
have played an important role in calculations of invariants of
3-manifolds, and in loop quantum gravity, where they provide a
gauge-invariant basis of states \cite{rovelli1,baez3}.  If we wish to
describe {\it space-time} by this type of method, we need, as we have
already remarked, an extension of the concept of spin networks.  An
alternative to the idea of relativistic spin networks is provided by
what has been called {\it spin foam} \cite{baez4}, because one can
think of a spin foam as a soap film connecting two spin networks at
different times.  \lq\lq Sums over surfaces" formulations of loop
quantum gravity have been given by Reisenberger and Rovelli
\cite{reisenberger2}, and
Iwaski \cite{iwasaki} has formulated the Ponzano-Regge model in
terms of surfaces.
Turaev and Viro \cite{turaev1} formulated their theory not only in
terms of a triangulation of the 3-manifold but also in terms of
simple 2-polyhedra forming a 2-complex embedded in the manifold, and
we can interpret
this second method as the first example of a spin foam model!  The
relationship between the evolution of spin networks and the approach
using triangulated manifolds has been explored and illuminated by
Markopoulou \cite{fotini1}.

The theory of spin foam is a way of formalising the calculation of the
partition function in BF theory by triangulating manifolds.  Recall
that a spin network is a graph with edges labelled by irreducible
representations and vertices by intertwiners.  Imagine moving such an
object through space, or rather space-time, so that it traces out a
2-dimensional surface, a generic slice through which would be a spin
network; this, heuristically, is what we mean by a spin foam.  It is a
2-complex, the faces of which are labelled
by irreducible representations and the edges by intertwiners.  The
dual triangulation of a manifold is an example of such an object. 

Baez \cite{baez1} has outlined how to calculate transition amplitudes
in BF theory using sums over spin foams, and the derivation of the
spin foam model from the classical action principle based on BF theory
has been discussed by Freidel and Krasnov \cite{freidel3}.  It has
already been
shown \cite{freidel4} that a particular type of spin network may be
evaluated as a Feynman graph, and the idea in the evaluation of spin
foam sums is to use Feynman's sum over histories approach, with BF
theory playing the r\^ole of the free theory and spin foams as
2-dimensional analogues of Feynman diagrams.  These techniques have
produced agreement with the lowest order terms in the known state sum
models \cite{freidel3}.  Markopoulou and Smolin \cite{fotini2} have
defined a model of the time evolution of spin networks based on local
causality rules, which are equivalent to those for spin foams.  

Recently Smolin \cite{smolin} has suggested a connection between evolving
spin networks, spin foam and such approaches related to loop quantum
gravity, and string theory, where there are clearly intuitive
similarities in the evolution of strings and membranes.  Any precise
equivalence still needs to be worked out, but Smolin's suggestion is
typical of recent ideas in which a number of apparently unrelated
approaches to quantum gravity seem at last to be coming together.

\section*{IV Area Regge calculus}

It seems that those attempts at formulating a theory of quantum
gravity in four dimensions described in the last section all need one
ingredient to be at all successful; this is the assignment of
labelling to the triangles instead of (or possibly as well as) the
edges.  (This fits in with work by Birmingham and Rakowski
\cite{birm3} who constructed state sum models based on $Z_p$ for
4-dimensional triangulated manifolds.  When the colourings from $Z_p$
were assigned only to the edges, the invariant depended only on the
3-dimensional boundary manifold, but when colourings were assigned
also to the triangles, the invariant depended on the 4-dimensional
structure.)   Even the spin foam description fits into this pattern
when  one
considers the triangulation to which it is dual.  By considering the
asymptotic value of the amplitude of a 4-simplex, we have seen that in
this case, it appears to be related to the path integral with the
Regge calculus action but with the triangle areas playing the most
important r\^ ole, rather than the edge lengths.

\subsection*{A Problems with the basic idea}

The idea that, in four dimensions, the triangle areas could be
regarded as the basic variables in a modified form of Regge
calculus was first suggested
by Rovelli \cite{rovelli2} and the possibility was discussed in some
detail in
\cite{barrett7}.  In this section, we shall consider the
advantages and disadvantages of the approach, and report on some
progress in understanding the relationship between the two types of
variable. 
 
A 4-simplex not only has ten edges, it also has ten triangles.  Thus
at first sight, the change from edge lengths to triangles areas as
basic variables looks very straightforward, but there are actually a
number of problems \cite{barrett7}.

Consider first a single 4-simplex.  It is simple to express the
triangle areas in terms of the edge lengths.  However, to express the
Regge action in terms of the new variables, we need to invert the
relationship between areas and edge lengths to be able to calculate
the deficit angles.  Unfortunately the Jacobian is singular in cases
where a number of triangles are right-angled and there is not
necessarily a unique set of edge lengths corresponding to a given set
of areas \cite{{barrett9},{tuckey2}}.  This means that, right from the
start, certain regions in
the space of edge lengths must be avoided.

Secondly, for a collection of 4-simplices joined together, there will
not in general be equal numbers of edges and triangles so there may be
ambiguity about which is the correct number of variables.

Thirdly, by considering two 4-simplices meeting on a tetrahedron with
all triangle areas assigned, we can envisage the following bizarre
situation.  Solve for the edge lengths of one of the 4-simplices in
terms of its triangle areas.  Repeat this for the other 4-simplex.  It
is possible that the edge lengths of the common tetrahedron will differ
according to the 4-simplex where the calculation was done (see
\cite{barrett7}
for an example).  Clearly there are difficulties in interpreting the
edge lengths as real physical quantities in the usual sense.

In this section, we shall now discuss possible theories in terms of
equations of motion and then investigate the dynamical content of area
Regge calculus by studying the weak-field expansion about a flat
background in terms of variations in the areas.

\subsection*{B Equations of motion}

The counting of degrees of freedom in a discrete theory is never
completely straightforward.  In a simplicial theory, the usual
argument is that in $n$ dimensions, an n-simplex has $n(n+1)/2$ edges,
which corresponds to the number of  independent degrees of freedom of
the metric tensor in $n$ dimensions.  If one thinks of these variables
as being at some chosen point in each simplex, the counting becomes
somewhat less clear when one realises that each of the edges is shared by
a number of other simplices, so the number of variables per point is
quite obscure.

Given this ambiguity, we can take two attitudes to the counting
problem in area Regge calculus.  Either we can take the areas as the
fundamental variables, worrying about the different numbers of edge
lengths only inasfar as we need them to calculate deficit angles or
volumes, or we can regard some of the areas as redundant variables and
aim to reduce their number to the number of edge lengths in the
simplicial complex.
 
In the theory where the areas are taken seriously as variables (which
is our principal interest here since we aim thereby to understand the
the models described as 4-dimensional generalisations of the
Turaev-Viro theory), we concentrate on the restricted class of metrics
where the Jacobian is non-singular.  Then the hyperdihedral angles
are well-defined and the Regge action may be written as 

\begin{equation}
{I_R(A_s)} =
{\sum _t A_t \epsilon_t(A_s)}
\end{equation}

where the sum is over triangles $t$ and $\epsilon_t$ is the deficit
angle at triangle $t$.  Variation of the action with respect to the
area $A_u$, use of the chain rule, an interchange of the orders of
summation and use of the Regge identity [1] leads to

\begin{equation}
{{\epsilon_u} = {0} \ \ \ \  for \  all \ u.}
\end{equation}

For details, see \cite{barrett7}.  Since all deficit angles vanish,
the space is
locally flat; the holonomy round any triangle is trivial.  This agrees
with Ooguri's state sum model for BF theory \cite{ooguri2}.  The
interpretation of this result is not obvious and the investigation of
such spaces using parallel transport is under way.

The other possibility, that of regarding some of the areas as
redundant variables, has been investigated by M\"akel\"a
\cite{makela1}.  Clearly in
order to recover the conventional view of simplicial gravity where the
edge lengths are real physical quantities, it is necessary to impose
the condition that a given edge has the same length in whichever
4-simplex that length is calculated.  This leads to a large number of
constraints: for each edge, there is a constraint for each pair of
4-simplices meeting there.  For a simplicial complex with $N_1$ edges
and $N_2$ triangles, a total of $N_2 - N_1$ of these constraints will
be independent, but it is not easy to give any general rule for
picking out which these are.  (An {\it ad hoc} rule has been formulated for
a particular model and it is likely that there is some group-theoretic
basis for the rule \cite{makela2}).  M\"akel\"a has shown that if the
variations of
the constraints are added in with Lagrange multipliers to the
variation of the Regge action expressed in area variables, then the
usual Regge calculus equations of motion are recovered.

\subsection*{C Dynamics}

Restricting our attention now to the area variable theory without
constraints, we investigate its dynamical content by performing a
weak field expansion about a flat background \cite{rocek4}.
This is in analogy with
the weak field expansion for edge length variables \cite{rocek2},
which we now
describe briefly.

In the original calculation, a 4-dimensional hypercubic lattice is
divided into simplices by drawing in various diagonals, giving fifteen
edges per vertex.  Small variations of the edge lengths about their
flat space values are made by setting

\begin{equation}
{l_i} = 
{l{_i}{^{(0)}} (1 + \delta _i)}
\end{equation}

with $\delta _i \ll 1$.  The second variation of the Regge action (the
first non-vanishing term) is evaluated as a quadratic expression in
the $\delta$'s, written as 

\begin{equation}
{\delta ^2 S} =
{\delta _i M_{ij} \delta_j},
\end{equation}

with $M_{ij}$ a sparse infinite dimensional matrix.  A Fourier
transform is then performed by relating $\delta$ in the $n$
direction and based at the lattice point $(i,j,k,l)$ steps in the
$(1,2,4,8)$ directions from the origin (see \cite{rocek2} for details
of the
binary notation) to the corresponding $\delta$ at the origin by 

\begin{equation}
{\delta{_n}{^{(i,j,k,l)}}} =
{\omega_1^i \omega _2^j \omega _4^k \omega _8^l \delta{_n}{^{(0)}}}
\end{equation}

with $\omega _\mu = exp(2 \pi i/n_\mu)$, where $n_\mu$ is the period
in the $\mu$-direction.  Acting on periodic modes, $M$ reduces to a
block diagonal matrix with 15x15 dimensional blocks, $M_\omega$.  This
matrix $M_\omega$ has four zero modes , corresponding to periodic
translations of points of the lattice, and a fifth zero mode
corresponding to periodic fluctuations of the hyperbody diagonal.
Block diagonalising $M_\omega$ decouples four further modes; they
enter without $\omega$'s  and so do not contribute to the dynamics at
all.  Their equations of motion constrain them to vanish.  We see from
this that an apparent mismatch in the number of components (fifteen
per vertex) is corrected by the dynamics of the theory, leaving ten
degrees of freedom per vertex, as would be expected from the continuum
theory.  (The zero modes correspond of course to gauge fluctuations.)

We now perform the analogous calculation with area variables.  In this
case, it is necessary to use a \lq\lq distorted" hypercubic lattice
because
the original one contains many right angles which lead to vanishing of
the Jacobian when transforming between areas and edge lengths.  This is
obtained by squeezing each unit hypercube along its hyperbody diagonal
until it has length $1$ in lattice units, like the edges originally
along the coordinate axes.  The face and body diagonals then all have
length $\surd (3/2)$.  Small variations of these edge lengths about
their flat space values are then made and the second variation of the
action within each 4-simplex calculated.  These variations in edge
lengths induce changes in the triangle areas represented by

\begin{equation}
{A_i} =
{A{_i}{^{(0)}} (1 + \Delta _i)}
\end{equation}

with $\Delta _i \ll 1$.  Within each 4-simplex, the expressions for
the $\Delta_i$'s in terms of the $\delta_i$'s are inverted (uniquely)
and the second variation of the action written in terms of the
$\Delta_i$'s.  Adding together the contributions from all 4-simplices
gives 
\begin{equation}
{\delta ^2 S} =
{\Delta_i N_{ij} \Delta_j},
\end{equation}

with $N_{ij}$ again a sparse infinite dimensional matrix.  A Fourier
transform is then performed as in the edge-length variable case, and
$N$ reduces to a block diagonal matrix with 50x50 dimensional blocks
$N_ \omega$ (note that there are 50 triangles based at each vertex).
The size of $N_ \omega$ makes it necessary to investigate the modes
numerically, and, somewhat contrary to our original expectations, it
turns out that the number of dynamical modes is exactly the same as in
the edge length case.  There are again four zero modes, corresponding
to periodic fluctuations of the lattice, and six further modes scaling
with $k^2$, where $k$ is the momentum in the Fourier transform.  The
remaining forty modes enter non-dynamically (they are massive and do
not scale with momentum) and are constrained to vanish by their
equations of motion.

Thus the theory with area variables is equivalent to the edge length
variable theory from the point of view of dynamical content.  This is
very encouraging and gives impetus to the search for the exact
correspondence between the variables in models like that of Barrett
and Crane, the variables of Regge calculus and ultimately the
variables of conventional general relativity.  That search continues.

\section*{Acknowledgements}

The authors thank John Barrett and Radu Ionicioiu for help
with the preparation of this paper.  The work was supported in part by
the UK Particle Physics and Astronomy Research Council.

\end{document}